# Flash Temperature and Force Measurements in Single Diamond Scratch Tests

Mansur Akbari [a*], Mikhail Kliuev [b*], Jens Boos [a], Konrad Wegener [a, b]

[a] inspire AG, ETH Zurich, Technoparkstrasse 1, 8005 Zurich, Switzerland
[b] Institute of Machine Tools and Manufacturing, ETH Zurich, Leonhardstrasse 21, 8092 Zurich, Switzerland

*These authors contributed equally and share the first authorship
Corresponding authors: akbari@alumni.ethz.ch, klyuev@iwf.mavt.ethz.ch

## Abstract

Analysis of the highest temperature in the machining processes, namely the flash temperature, helps to understand the physics of the process, to improve cutting tool geometry and to achieve high performance machining. In the present work, the interaction between cutting grain and workpiece material in grinding process is analyzed. Single diamonds are considered for machining, which operates in comparison to other measurements in the range of grinding speed. The highest temperature in the grain-material interaction and cutting forces are measured. In order to measure the flash temperature an innovative method to measure and analyze the temperature through the diamond grain in the cutting zone by two-color pyrometer are proposed. Furthermore, cutting forces are measured simultaneously. In order to measure the temperature in the cutting zone, an accurate connection between diamond and pyrometer fiber is required. Thus, the diamond tool holders are manufactured by electrical discharge machining (EDM)-milling in deionized water. A 0.5 mm diameter hole is drilled in each holder, to connect the diamond precisely to the pyrometer fiber. Machining processes are performed with 30 μm depth of cut, cutting length of 20 mm and cutting speed of 65 m/s on Ti6Al4V. The cutting tool is fixed and the shape of the rotating workpiece is optimized. The diamond holder with the specific shape is designed and manufactured. Quasi-static, dynamic, modal and harmonic response analyses are performed in order to reduce vibrations and chattering. The measured flash temperature is 1380 °C and cutting normal, tangential and axial forces are measured.

## 1. Introduction

Flash temperature is the highest temperature in machining process, which plays the key role in high performance machining. Flash temperature assessment is important in high performance machining of hard-to-cut materials, used in various industries such as aerospace and biomedical applications. Some applications of flash temperature and cutting forces analyses are studying its influence on tool wear, surface integrity of the machined part and the final cost of manufacturing. In dynamic process of machining, the flash temperature occurs in the interaction of the cutting tool with the workpiece, directly in the cutting zone. Thus, measurement of flash temperature with conventional methods, such as IR-thermography has poor accessibility to the cutting edge, where the temperature is the highest. High localized flash temperature, as it is reported by [1], is one of the sources of thermomechanical failure and tool-wear. Measurement of the flash temperature in machining process helps to validate machining simulations and as a result, it helps to optimize the grinding process and the design of the machining tools. One of the challenges in machining titanium alloys, i.e. Ti6Al4V, is its low heat conductivity, where the heat does not transfer to the chip or inside the workpiece and concentrates in the contact area during machining operations. In high energy-consuming processes like grinding, where the chips are not able to take away enough heat, it is important to dissipate the heat through cooling.

Three methods are commonly used for measuring cutting temperature: resistance thermometer, thermocouple and radiation thermometers, but only methods based on the IR-thermography because of the flexibility of the measurement setup are widely used for temperature measurements in grinding process. According to [2], infrared radiation is in a wavelength range between 0.78 μm and 1000 μm. In single-color pyrometers, as mentioned in [3], easier construction and lower signal noise make higher temperature resolution possible. In order to avoid emissivity dependency on temperature, two-color pyrometers, which measure at two different wavelengths, are often used.

As it was suggested in [4], infrared radiation investigation, measured by infrared (IR) camera, gives valuable insight into the process of grinding. IR thermo-camera recorded the radiation in grinding at 30 m/s. This analysis showed that grinding wheel could not cool completely during one revolution. Excessive heat exposure and wear of the grains lead to noticeable change in radiation. In order to understand the physics of grinding process, as shown in [5], one can set sights on a single grain and analyze the grain-material interaction, grinding forces and temperature. However, measurement of flash temperature in the cutting zone is currently impossible with infrared camera. In addition, with infrared camera, there are also challenges in misinterpreting the results due to not knowing the emissivity of the object. In the present work, these challenges are tackled in a single grain scratch test with proposing a new measurement setup. With the help of numerical analysis, special tool holder is designed to facilitate the measurement. For the force measurement, piezoelectric sensors are used.

The experiment setup with a single-color pyrometer described in [6] indicated that in grinding of 0.55 % carbon steel, with $Al_2O_3$ grinding wheel, peripheral wheel speed of 1713 m/min, workpiece speed of 10 m/min and 20 μm depth of cut, the temperatures of 1400 °C are still held by a few grains even 25 ms after material contact. Optical fibers were placed directly beyond the contact zone to the workpiece orthogonal to the wheel to transfer IR rays to a single-color pyrometer. In another similar setup in [7] three different grinding wheels with $Al_2O_3$, CBN and diamond grains were studied. It was observed that the coolant could not reach the active cutting point where the material separates, and it was impossible to avoid heat generation. In addition, it was found that the temperature of cutting grains hardly varies due to cutting depth and workpiece speed but with change of wheel velocities.

Two-color pyrometer is widely used for temperature measures in the cutting zone. The research in [8, 9] reported the dependency of the cutting speed on cutting temperature. Single diamonds and real grinding wheels at approximately 30 m/s grinding speed are analyzed in [9]. The workpieces were made of Nickel-zinc and Manganese-zinc ferrites ceramics and carbon steel with embedded IR sensors. Additionally, out-of-roundness of the grinding wheel has leaded to spikes in temperature. The CBN grinding wheel in the test setup by [10], has operated with cutting speed of 58 m/s. Three thermal measurement methods were examined. Two-color pyrometer and flash temperature peaks were measured. According to [11], those temperature spikes originate from shear planes in grinding chips and approach the melting point of the workpiece material. On the other hand, as mentioned in [12], the energy and consequently heat flux is proportional to the grinding speed.

The influence of the distance between optical- fiber and heat source was analysed in [13]. In their study it is shown that in the range from 0.2mm to 2mm the measured emissivity of the blackbody with the temperature, fixed at 650 °C remains the same. Therefore, they have proved that in their case study, the change of the distance in this range did not have a significant influence on measured temperature. An infrared two-color pyrometer with low-noise photodetectors and high-gain transimpedance amplifiers was tested in [14] for temperature measurement. The pyrometer was able to measure temperature in localized areas at a high speed with the same gain factor in the range of 170°C to 530°C. In another study in [15], a new temperature measurement method of flat glass edge during grinding, which used constantan (a copper–nickel alloy) and copper strips on both sides of the glass plates is shown. The method helps to adjust the machining parameters by the glass temperature measure. For machining parameters study in grinding [16] used a two-dimensional sliding table, the focus of infrared thermometer to adjust to the small hole in the workpiece in order to accurately measure grinding temperature. IR thermography was also used in [17] to monitor grinding conditions. The measurements were facilitated to define the right strategy in the grinding of Metal Matrix Composites (MMCs). A conical $Al_2O_3$ cutting tool was used in [18] for grinding 0.55% carbon steel and an optical fiber was connected to a two-color pyrometer. Depth of cut of 30 μm and cutting speed of 100-2300 m/min are used and maximum temperature of 1400 °C is measured. As mentioned in [19], the temperature depends on the thermal diffusivity of the metal workpiece. The higher is the thermal diffusivity the lower is the maximum temperature. Three teeth of carbide were mounted with single-color pyrometer in processing steel in an end milling cutter test setup in [19]. A maximum quasi-stationary temperature of 1400 °C was measured. The dependency of temperature on thermal material properties and cutting speed was verified experimentally and with finite element method (FEM) in [20]. Ueda et al. also stated that pyrometers were the best option for temperature measurements in single grains, since the grains are too small. In their experimental setup, the diamond tip is attached to the tool shank, which has a small gap at its top to insert the optical fiber of the pyrometer, which cannot measure the flash temperature in the contact region. The setup contained single diamond arrangement in turning at 10.33 m/s using copper and aluminum as workpieces. To improve tool life and integrity real time temperature measurements in real machining are requested. The dependency of temperature on the cutting speed in micro-grinding and micro-milling was also shown by [21].

Multi-color pyrometers with three or more channels also exist but they are not widely used. Coates [22] calculated the temperature error for multi-color pyrometers and recognized an intense error increase by increasing the number of channels. The same problem was investigated by [23] for two-, three- and four-color pyrometers at different metallic surfaces. They also did not figure out any remarkable benefit of concept with more than two channels. It was shown in the literature that the two-color pyrometer is the most precise method for measurements of the temperature in the cutting zone. However, measurement of flash temperature in the cutting zone for the grinding tools, in addition to measuring the cutting forces simultaneously are not found in the literature.

The present work describes an innovative measurement setup with the two-color pyrometer and piezoelectric sensor to determine both temperature and force simultaneously during the machining process. The innovative design allows the pyrometer to measure the temperature directly in the cutting zone by detecting IR radiation through the transparent diamond cutting-cool. The details of the experimental setup are presented in section 2. In order to reduce the vibrations for the rotary tool, in section 2.3, the shape of the rotary tool is optimized, in addition to performing modal analysis and harmonic response assessments. The assessments of the force and temperature results are presented in section 3. Subsequently, a conclusion is presented in section 4.

## 2. Experimental setup

### 2.1. Two-color pyrometer

Two-color pyrometer (FIRE-3, RWTH, Aachen, Germany) is used in this work for flash temperature measurements. The emissivity of light during cutting process varies with wavelength, temperature, direction, material properties and surface condition. In order to lower complexity of emissivity in the measurements, it can be idealized as a gray body. The gray body assumption declares that the body's emissivity is independent of wavelength and direction; more details can be found in [24]. Measurements made by Davies in [25] displayed that gray body assumption is not valid for most surfaces and wavelength. Therefore, it is necessary to select a measurement method that is not based on gray body assumption for flash temperature measurements in micromachining processes. To overcome the problem of the unknown accurate values of emissivities in the cutting zone in measurement of the temperatures for different surface roughnesses, temperatures and change of the material properties, two-color pyrometers are used. Furthermore, access to the machining zone and flash temperature measurement in the cutting zone are complicated. In the present work, this goal is achieved by facilitating a two-color pyrometer and special setup to look through the diamond grain and measure the flash temperature in the cutting zone. The details of this setup are explained in more details in next sections.

As mentioned in [24], two-color pyrometers are ratio thermometers that collect radiation at two different wavelengths by forming the ratio between the two input signals. The two-color pyrometers are based on ratio of energy measured at two wavelengths in order to measure the temperature. These two wavelength filters are laid one on top of the other. One wavelength is a broad wave band and the other wavelength is a narrow wave band that is a subset of the broader band. As explained in [26], the two wavelengths that the temperature is evaluated are 1.7 and 2.0 µm. An error will only occur, if the emissivities are dissimilar at the two measured wavelengths. If emissivity is assumed at the two wavelengths, the temperature is called ratio temperature $T_R$, and it relates to the measured temperature by pyrometer. As indicated in [3], the surface temperature $T$ can be calculated from the measured temperature $T_R$ from:

$$T = T_R - \frac{C_2(\lambda_2^{-1} - \lambda_1^{-1})}{ln(\varepsilon_1/\varepsilon_2)} \tag{1}$$

The measurement error contains emissivity, $\varepsilon$, wavelength, $\lambda$, and radiation constant, $C_2$. If the measured body is a gray body, equation (1) shows that the measured temperature and real temperature are similar. The equation (1) indicates an increasing error, if the gap between $\lambda_1$ and $\lambda_2$ is small.

### 2.2. Measurement setup

The machining-center that is used for performing the experiments is PRÄZOPLAN 300. In the present work, 3D piezoelectric dynamometer MicroDyn as introduced in [27], is used to measure the three components of cutting force. The lowest frequency of this dynamometer is in the range of 15000 Hz, therefore for this study and the measurements up to 5000 Hz, is applicable. The experimental test setup is illustrated in Fig. 1. A steel spacer is mounted onto the machine table and is assembled with MicroDyn sensor, tool holder and diamond tool. Aluminum is chosen as material for the tool holder because of the low density and, therefore, the low weight of the tool holder assembly. Any mass mounted onto the dynamometer lowers its eigenfrequency and restricts the measurement range.

The Ti6Al4V disc is an elliptical disc with straight flanks, as shown in Fig. 1. It is possible to mill the flanks in future to get the shape of the optimized disc. The pyrometer fiber is connected to the diamond and is able to measure the temperature through the diamond, directly at the cutting zone. For not overlapping the scratches at the workpiece flanks, the spindle moves upwards along the Z-axis with $V_z$. For the small diamonds, the experiment is made at a cutting speed of $V_c = 65$m/s.

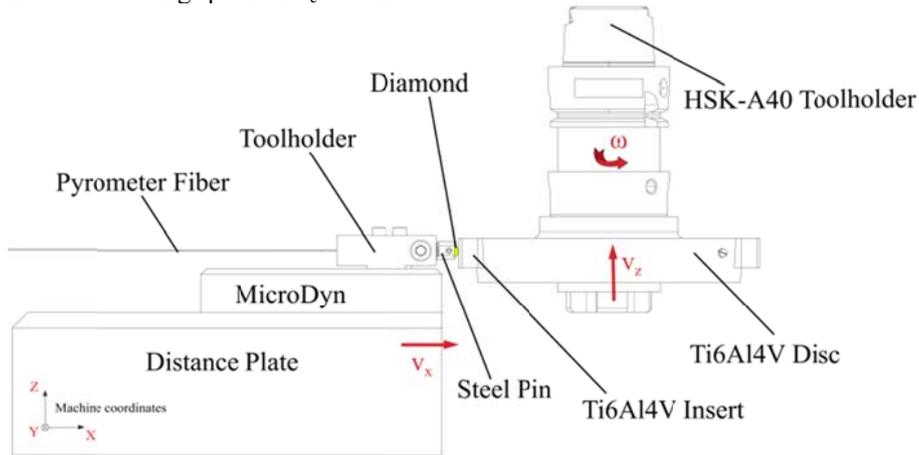

Fig. 1: Schematic of the experimental setup

A turning tool, as it is shown in Fig. 1, is placed next to the MicroDyn dynamometer to remove the scratches on the insert flanks after analyzing. The small diamonds are aligned with digital handheld microscope of Dino-Lite Premier AM7013MZT (AnMo Electronics Corporation, New Taipei, Taiwan) for a vertical cutting edge as shown in Fig. 2.

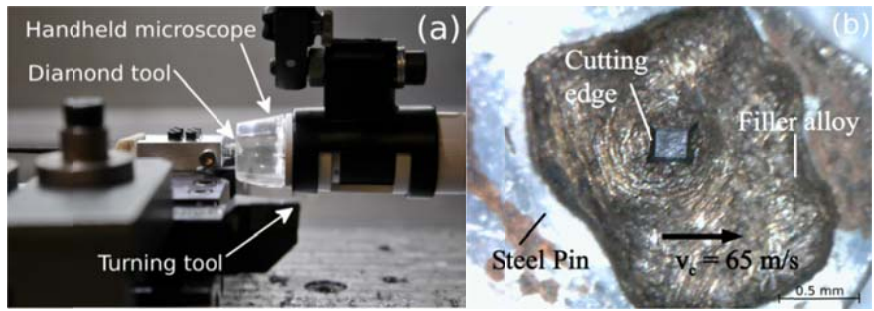

Fig. 2: (a) Digital handheld microscope is used for adjusting the orientation of the diamond grain before performing the experiment. As shown in [28], in order to have repeatable results, it is important to have correct orientation of diamonds in brazed diamond tools. Therefore, in (b), the component of brazed diamond grain, orientation of the grain, cutting edge and cutting velocity direction are shown.

## 2.3. Design and optimization of the rotary workpiece holder

For measuring the flash temperature in the cutting zone, the fiber of the pyrometer needs to be located close to the cutting zone or measure through a transparent object. In the present work, the temperature is measured through the diamond via EDM drilled hole inside the tool. Since the fiber of the two-color pyrometer is passed through the tool and cannot be rotated with it, in the present work, the workpiece rotates and not the tool. One of the common problems that can happen between the current setup and normal machining process is the impact of the tool and workpiece in entrance of the tool to the workpiece. To solve this issue, as mentioned in [29], in the edge of the rotating workpiece that impacts happens a "waterfall" form ,which is a half-parabolic shape, are machined in the experiments. The limitation in the spindle load requires weight and inertia reduction of the rotating part. However, for high rotational speeds and short contact lengths it is important, that the lowest eigenfrequency of the setup is minimized. The CAD model of rotating workpiece after optimization is displayed in Fig. 3. The optimized tool has a conical shape. Compared to other parameter combinations, this design lets enough space for the insert screws, which are important to guarantee the inserts fixation. The holes for these screws are also considered in the followed FEM analyses. For accurate FEM simulations, the assembly is analyzed with ANSYS® software.

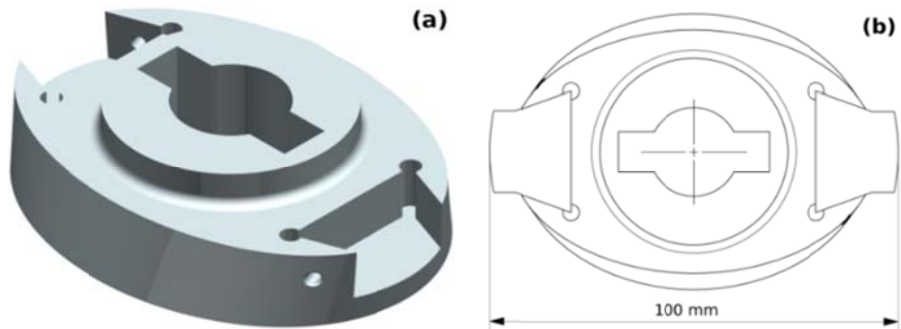

Fig. 3: 3D model of the workpiece holder with two inserts. (a): The workpiece holder; (b): Assembled workpiece holder with its two workpiece inserts

To analyze the response of the structure in both quasi-static and dynamic analysis, structural analysis under centrifugal loads is performed with applying the rotational velocity to the tool. The structural analysis gives information about the total deformation and equivalent stress, caused by centrifugal force as displayed in Fig. 4. The simulation results in Fig. 4a, indicates that the total deformation of the whole assembly and the workpiece is negligible. Furthermore, the von-Mises stress results in Fig. 4b proves insignificant stresses in the assembly with respect to the yield point of the titanium alloy.

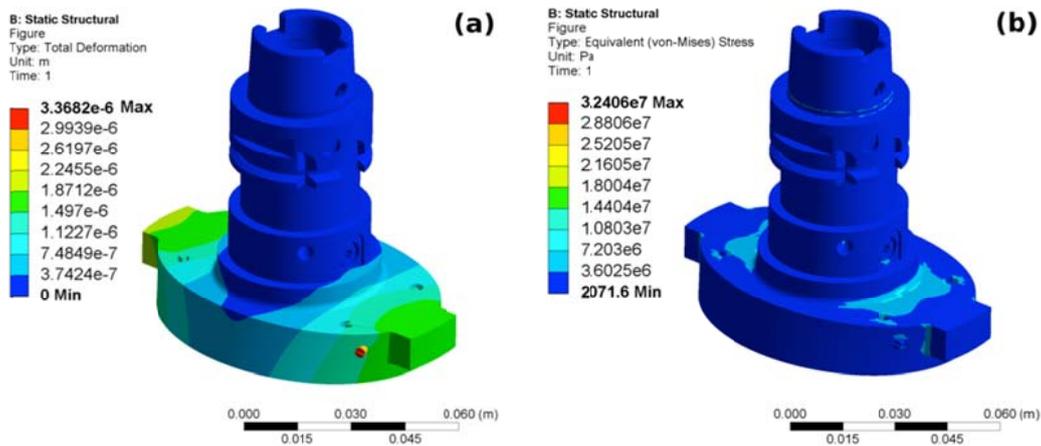

Fig. 4: (a) The total deformation and equivalent stress are demonstrated. (b) von-Mises stress in the assembly shows that the maximum total deformation of insert flanks is 1.92 μm. The critical value of the von-Mises stress in the workpiece holder is 25.98 MPa; the maximum stress is located inside in the workpiece-holder.

Modal analyses are also performed to get information about eigenfrequencies and their appropriate mode shapes. All boundary conditions, materials and mesh are imported from quasi-static analysis. The rotational velocity as well as damping is not considered in the modal analysis.

The first eight eigenmodes of the system are calculated and the eigenfrequencies are listed in Table 1. These frequencies are able to lead to resonance. If an excitation frequency from the cutting process is close to one of the calculated eigenfrequencies the system can generate resonance.

Table 1: Eigenfrequencies of the rotary disc, mounted on a HSK interface. Eigen modes have been calculated and are shown in Fig. 6

| Mode | Frequency [Hz] |
|------|----------------|
| 1    | 1904.9         |
| 2    | 1957.8         |
| 3    | 3720.8         |
| 4    | 7285.3         |
| 5    | 8527.5         |
| 6    | 9553.1         |
| 7    | 12548          |
| 8    | 15351          |

As explained in [30, 31], in modal analysis and other eigenvalue-based analyses, the results show the deformed shape scaled by an arbitrary factor. Therefore, in this study, modal analysis explains qualitative state of deformation within each mode and not the real physical deformation. Not all modes are identically critical. To find out which eigenmodes are the most critical ones, harmonic response study is performed. The cutting forces are considered as excitation forces to get the real amplitudes for each eigenfrequency.

For solving equation (2), mode superposition is considered. This method takes mode shapes and eigenfrequencies from the previous modal analysis to compute the frequency response of the system under dynamical applied loads. The solver transforms the basic equation from nodal into modal form and solves for deformation. The procedure in more details is described in [32].

$$( [K] - \Omega^2 [M] + i\Omega[C] ) ( \{u_1\} + i\{u_2\} ) = \{F_1\} + i\{F_2\}$$

(2)

where [C] is the damping matrix, $\Omega$ is the imposed circular frequency, $\{u_1\}$ is the real displacement vector, $\{u_2\}$ is the imaginary displacement vector, $\{F_1\}$ is the real force vector and $\{F_2\}$ is the imaginary force vector. Prior to cutting experiments, harmonic response analysis is performed to calculate the permissible process parameters in the experiment. According to the general grinding theory and as mentioned in [11, 33], a normal force is about three times higher than the tangential force and much higher than the axial force. If this assumption is not valid, the main potential impact is to the experimental setup i.e. vibration of the rotating workpiece, which could cause errors and noise during force measurements. Analysis of eigenfrequencies is used for right selection of rotation velocity. Since no significant vibration is detected during experiments, correct rotation velocity was selected. The location of applied forces, as shown in Fig. 5, is considered where the initial impact takes place and causes vibrations in the system. The applied forces in normal, tangential and axial directions are 10 N, 3 N and 1 N respectively.

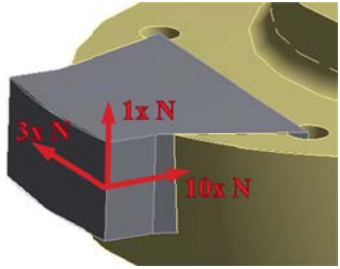

Fig. 5: Assumed cutting forces on the workpiece for eigenfrequencies calculation are depicted.

To eliminate large amplitude peaks, damping factor of 0.01 is assumed. Damping factor consists of material and structural damping, whereof the last one is more important. Structural damping coefficient depends on the system geometry. Therefore, an assumption is made which is sufficient for evaluating the most critical eigenmode qualitative.

The result of the harmonic response study is plotted in Fig. 6 and the eigenmodes from the modal analysis are marked. The amplitude is similar to the deformation. It can be seen in Fig. 6, that the first eigenmode (at 1905 Hz) is the most critical one. Even if the damping factor or the force values are changed, the first eigenmode always leads to the highest amplitude.

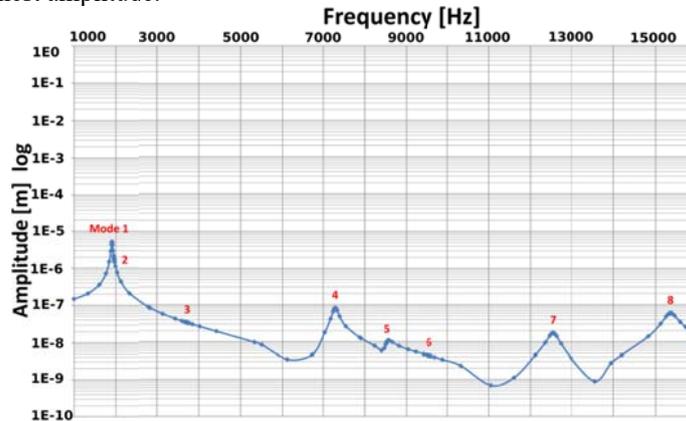

Fig. 6: The first eight eigenfrequency of the rotating workpiece for the spatial forces shown in Fig. 5 are depicted in red colors. The values of the eigenfrequencies are given in Table 1.

The frequency response study illustrates that the first eigenfrequency should not be excited by process frequencies. The magnitude of excitation spectra is small above the low pass filter frequency of the charge amplifier. Furthermore, the impact frequency is the number of diamond engagements at the insert leading edge per second. For one insert this leads to an excitation frequency of:

11460 RPM / 60 = 191 Hz    (3)

Every multiple of 191 Hz is critical, if it is near an eigenfrequency. As mentioned above the first eigenfrequency at 1905 Hz is the most critical one of the system. Since, 1905 Hz is nearly ten times the excitation frequency, thus resonance can occur. It is not important whether the most critical eigenmode is at a higher or lower frequency. It is important to know at which frequency it is critical. Thereafter the machine revolution can be set, that the critical eigenfrequency is not exceeded. For this reason, the cutting speed is increased from 60 m/s to 65 m/s and consequently the excitation changes to 207 Hz. For this purpose, the first eigenfrequency is in the middle between the ninth and tenth multiple of the excitation.

EDM shaped diamond holder is used for precise diamond positioning and robust fixation, an accurate diamond holder is required. The diamond holder is a cylindrical pin made of stainless steel X2CrNiMo18-14-3, which is screwed into the tool holder of the dynamometer. The diamond is mechanically clamped in the holder to avoid displacement. Monoplate MXP L4015 diamonds from Element Six Company are used for the flash temperature evaluation in micro machining of TiAl6V4.

In order to pass the fiber of the pyrometer through the tool holder and measure through the diamond, accurate and deep holes are machined in the tool holder. EDM-drilling and milling are challenging processes due to high hole-aspect ratio (0.5 mm hole with 15 mm depth in this case study). This high hole-aspect ratio can lead to an electrode break and can cause inaccuracy of the shape. Furthermore, imprecise drilling can cause the contact between diamond and electrode can cause graphitization on the contact diamond surface, based on [34], and may interfere the temperature measurement. Two EDM processes can be used for cavity manufacturing application. The first process is die-sinking EDM process, which is usually used for the complex cavity manufacturing, but it also requires manufacturing the complex shape of the electrode. Since all diamonds requires different size of holder, several new electrode needs to be manufactured for the die-sinking. However, EDM-milling, which is the second suitable EDM process, has other advantages, which is used in the present study. By using EDM-milling, it is possible to apply standard cylindrical electrode for different cavities manufacturing.

A hole with diameter of 0.5 mm is drilled in the center of the holder in order to connect the fiber of pyrometer. The holder is made on CNC EDM-drilling machine (AgieCharmilles DRILL 300, GF Machining Solutions, Switzerland).

The pin of brazed diamond tool is manufactured by the layer-by-layer milling. The electrode has the main vector of milling movement, which is used for cavity creation. The electrode wear compensation vector, is used for the frontal wear compensation and electrode rotation movement applied for uniform lateral wear. Every machine pass, erodes a layer of material on the surface of workpiece and produces the required geometry. The wear compensation vector depends on the technology and workpiece material. Layer-by-layer EDM-milling is explained by Yu in [35] in more details.

The erosion current, $I_d$, is set to 8 A and open voltage, $U_{op}$, is set to 100 V. As mentioned in [36], the layer thickness per one pass cannot be higher than a half of electrode radius, which is considered in the present work. The diameter of the pyrometer fiber is 0.5 mm. Therefore, the diameter of the electrode for drilling and milling, which is used in this work has the same diameter as the fiber. The maximum thickness of each layer in the EDM milling process will lead to lateral erosion. In addition, the wear of the electrode causes instability in the process. As a result, the layer thickness, which is considered in the present work, is maximum 0.25 mm.

## 2.4. Diamond tools

Two different sizes of diamonds are tested in single grain scratch tests. The big diamonds have nominal length of 4 mm and the small ones are less than 850 µm. In order to test the experimental concept initially, the big diamonds are used. Then, for final experiments, the brazed diamond tools are used. That is because connecting the pyrometer and the process observation is easier in large scale. However, for the experiments with big diamonds, the pyrometer will not observe the cutting edge but the clearance surface. The small diamond grain represents a real grit size in coarse grinding wheels.

Fig. 7 shows a 3D image of the EDM shaped cavity for the diamond grain, the drilled hole for the fiber and the diamond connection. A M2 screw from one side clamps the diamond. The thread hole for the M2 screw and the screw are not displayed in Fig. 7.

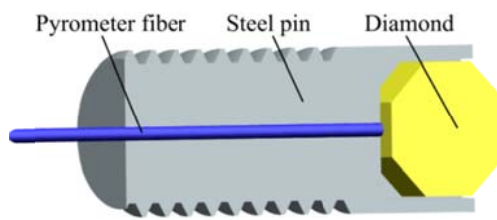

Fig. 7: Section view of the tool-holder, illustrating the pyrometer connection with a big diamond; the other side of the fiber, which is free in the figure is guided to the two-color pyrometer.

Uncoated truncated-octahedron single crystalline synthetic diamonds (SDB™ 1125 2025, Element Six e6™, ServSix GmbH, Karlstein, Germany) are used in the brazed diamond tools. According to Element Six Company, these diamonds have high strength in high temperature and suitable for operating from 900 °C to 1100 °C. Due to the small size of these diamonds, they are brazed on a round pin of austenitic stainless steel X2CrNiMo18-14-3. The brazing process of small diamonds takes place in a high-vacuum furnace. The steel pins and the diamonds are positioned in the furnace with an additional small weight to press the diamond against the steel pin. This concept hinders the diamond grain from floating during brazing. As mentioned in [5], Cusil ABA is used for brazing of diamonds. The complete description of the brazing process is given by Buhl in [37].

The drilled hole from the bottom of the pin to the diamond grain is necessary to connect the fiber of pyrometer directly to the diamond. The pyrometer collects the emitted radiation from the cutting edge, which passes through the diamond. Schematic of the namely big diamonds in this work, Monoplate MXP L4015 diamond, are shown in Fig. 8, where E equals 4 mm is the normal length of the diamond, H is the normal thickness of the diamond equal 3 mm and L is the edge length of inscribed square. Values L and E are significantly different from one diamond to the other, but the value L have to be at least 75 % from value E, and the normal thickness H is always 3 mm. However, because of differences in edge length of inscribed square, L dimension in Fig. 8, the diamond holder should be also different from one diamond to the other.

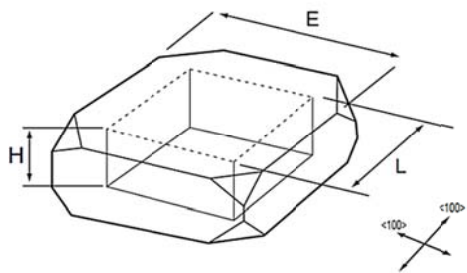

Fig. 8: Schematic view and crystallographic orientation of Monoplates MXP L4015; the diamond grain has E nominal length of 4 mm, T nominal thickness of 1.5 mm and L edge length of 3 mm.

The big diamonds are mechanically clamped and not brazed like the small diamonds. A round pin of stainless steel is used as diamond holder with a M6 thread. The pin is screwed into the tool holder and thereby fixed onto the dynamometer. An EDM drilled hole along the centerline of the round pin is made to connect the pyrometer to the diamond. The procedure to drill the hole is the same as in case of the small diamonds.

## 3. Measurement results

Monoplate MXP L4015 diamonds are used to test the experimental setup. Cutting speed of 32 m/s is set for the first run. The measured temperature is above 1200 °C. After performing the cutting experiments, the diamond tool, which is used in the experiment, is imaged by scanning electron microscope (SEM) performed by Quanta™ 200 F FEI and the properties of the remained material on the diamond edge are analyzed by energy-dispersive X-ray spectroscopy (EDX). As it is shown in Fig. 9, there is a layer on tip of the diamond edge. For the temperature measurements, it needs to be analyzed if it is graphitized layer or it is buildup-edge.

It is proved in [38-40] that when diamonds are heated to a high temperature, the changes that take place depend markedly on the environment around the diamond. If oxygen or other active agents is present, a black coating can form on the surface of diamond above about 600°C. This is not true graphitization which involves the transition of diamond to graphite without the aid of external agents. As shown in [41], if diamonds are heated in inert atmosphere the onset of graphitisation can be detected at 1500 °C. In other studies, such as [34], it is reported that surface graphitization of diamond take place at temperatures above 1800 °C, where the bulk transformation of diamond into graphite takes place.

In the present study, after performing the EDX in different regions of the cutting edge no graphitization is observed. Furthermore, as it is shown in Fig. 9, the layer on top of the cutting edge is the buildup-edge. The diamond that is shown in Fig. 9 is damaged after eight cutting experiments.

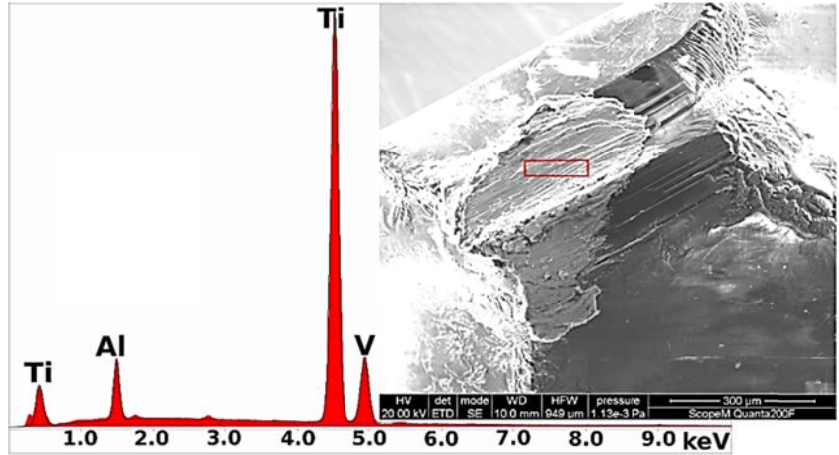

Fig. 9: Big diamond cutting edge after eight scratches at $V_c$ = 32 m/s. The diamond is imaged by SEM and EDX. The red rectangle in the SEM image shows the location of performing the EDX measurement and proves formation of buildup-edge. The EDX result in that rectangular region shows the high percentage of Ti in that region.

After testing the setup with big diamonds, the following parameters in Table 2 are considered for the small diamonds:

Table 2: Testing parameters for small diamonds

| Parameter | |
| --- | --- |
| Spindle speed $n$ | 12171 RPM |
| Cutting speed $V_c$ | 65 m/s |
| Speed Z-axis $V_z$ | 0.22 m/s |
| Depth of cut | 15 µm … 35 µm |
| Pyrometer amplification gain | $10^7$ |
| Pyrometer sampling | 500 kHz |

Due to uncertainties in defining exact 30 µm depth of cut, as shown in Fig. 10, scratches are started from 15 µm depth of cut of increased to more than 30 µm. With help of small feed rate in X-direction, the scratches in different depth of cuts are made. Experimental conditions are shown in

Table 3:

Table 3: Experimental conditions

| Parameter | |
| --- | --- |
| Number of scratches | 18 |
| Length per scratch | 20.1 mm |
| Time per scratch | 0.0003 s |
| Total process time | 0.045 s |
| Depth of cut | From 15 µm to 35 µm |

### 3.1. Analysis of scratches

A tactile profilometer of Form Talysurf™ Series 2 (Taylor Hobson™, Leicester, UK) is used to measure the profile of the scratches after the cutting process. As explained in more detail in [42] a diamond stylus with general characteristics of 60º cone angle and $2^{\pm 0.5}$ µm tip radius is used. The scratches for the two inserts are depicted in Fig. 10. The points in x axis of Fig. 10, as mentioned in [43], are referring to the measurements points, which represents the real profile of the scratch. Before start of the main experiment and performing the scratched on the surface of the workpieces, the surface of the workpieces are machined to have exact distance from the center of the tool-holder with considering the dynamics of the machining system. Therefore, after performing the scratch test, the scratches are measured in one location. It is shown in Fig. 10 that the depths of cuts do not increase continuously for scratches. A possible explanation is that the aluminum tool holder is too soft and therefore the diamond grain is pressed away from the insert. If this phenomenon occurs, the force in normal direction cannot increase and thus the depth of cut remains at an almost constant value. Further, it is seen that the scratches on the second insert go deeper. Since the complete system of the tool-holder and workpiece is dynamically balanced in addition to testing of the inclination and convexity of the work piece, not constantly increasing of the depth of cuts, can be also interpreted as effect of remaining dynamic unbalance.

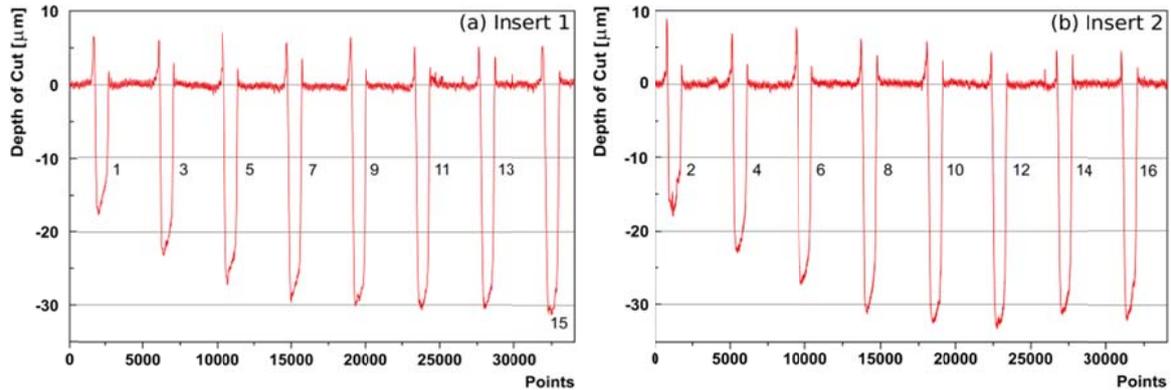

Fig. 10: Depths of the scratches on two inserts are shown. The points in x axis are referring to the measurements points that represent the real profile. It is visible that depth of cut stops increasing, after the eighth scratch. A possible explanation for that fact could be a too flexible tool-holder, made of aluminum, which is considered for its low weight on the dynamometer. The burrs on lateral regions of each scratch are due to plastic deformation from the cutting process.

### 3.2. Temperature analysis

As shown in Fig. 10b, the 8[th] scratch has 30 μm depth of cut. The temperature of this specific scratch is plotted in Fig. 11, which shows the temperature increase from 1000 °C to 1380 °C during the fast scratching process of 0.0003 s at 30 μs depth of cut. It can be seen that the pyrometer can measure such high speeds and temperatures. The signals oscillation ends exactly at the inlet flank of the inserts and starts again after the exit. The influence of different shapes and orientation of the diamonds, different depth of cuts, different speeds and different workpiece materials will be investigated in future researches.

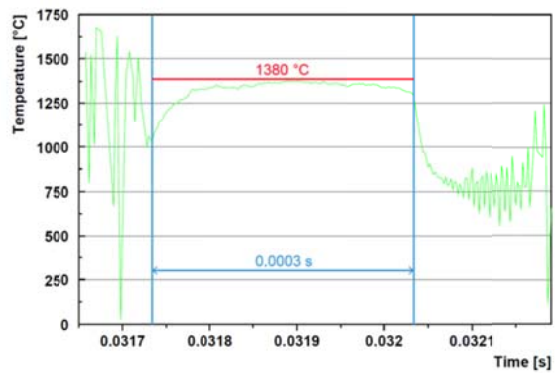

Fig. 11: The measured temperature for the 8[th] scratch with 30 μm depth of cut. The scratching time is indicated by the two blue vertical lines. As expected, the temperature increases until reaching the steady state. As shown, the plotted temperature curve has a visible starting and ending points; therefore, it can be concluded that the used pyrometer is fast enough to detect the high machining speeds.

### 3.3. Experimental setup and temperature analysis verification

An additional, thermal measurement of a simple defined heat source is done with the pyrometer (with and without the proposed innovative setup in this study) in addition to measurements with an IR thermo-camera. The experiment setup is shown in Fig. 12. As shown in Fig. 12, the setup contains a defined heat source from a soldering instrument (Weller, WES51 Analog Soldering Station, US) which is clamped and measured simultaneously with a two-color pyrometer and IR thermo-camera. The soldering tool as mentioned in [44] has uncertainty of ±6 ºC, however in reality it shows more uncertainty due to heating-cooling cycles . The two color pyrometer has the uncertainty of ±1 ºC [26]. As explained in [5] emissivity is a function of wavelength, surface roughness, direction and temperature, therefore it is needed to have a reference body with known emissivity to ensure that the correct output can be displayed during post processing. Thus, a black color spray which is matte from MOOD (JUMBO-MOOD, Ofenrohr-Lack, Switzerland) for making approximate black body reference with IR thermo-camera is sprayed on the soldering tip´s surface. The compared thermal results in Table 4, shows the same measured value by pyrometer at defined heat source of 350 ºC and through the diamond for low temperatures. However, for higher temperature measurements through the diamond, more noises are detected. Therefore, the calculated standard deviation of the thermal measurement in high temperatures with innovative setup is 28 ºC.

Table 4: The results of temperature analysis verification

| Heat source temperature | Pyrometer | Pyrometer in the experimental setup | IR-camera |
| --- | --- | --- | --- |
| 350 ºC | 390 ºC | 390 ºC | 309 ºC |
| 400 ºC | 425 ºC | 447 ºC | 353 ºC |
| 450 ºC | 457 ºC | 475 ºC | 408 ºC |

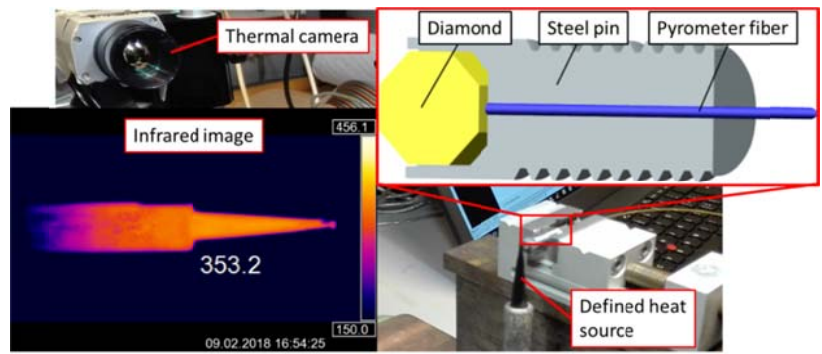

Fig. 12: Measurements verification test, measurements are simultaneously made by thermo-camera and by the experimental setup, where the defined heat source is observed through the diamond.

## 3.4. Force analysis

The forces should be at constant value after the impulse peak, because of a constant depth of cut over the scratch. However, an accurate force measurement for such extremely high dynamical process is hard to measure because of the oscillation aspects. The inlet flank of the inserts does not have a chamfer or a curved shape to start the cutting smoothly. Therefore, there is a high impulse when the diamond contacts with insert and can be treated as a Dirac pulse. The maximum force peaks are listed in Table 5.

Table 5: The results of the steady state machining forces for 30 µm depth of cut are shown. The error bar are derived from variation of results

| Parameter | |
| --- | --- |
| Normal | 115[86] N |
| Tangential | 141[87] N |
| Axial | 132[86] N |

The forces for all directions during the whole cutting process are depicted in Fig. 13. The oscillation problem especially in tangential direction is shown in Fig. 13. Normal and axial force have nearly the same values.

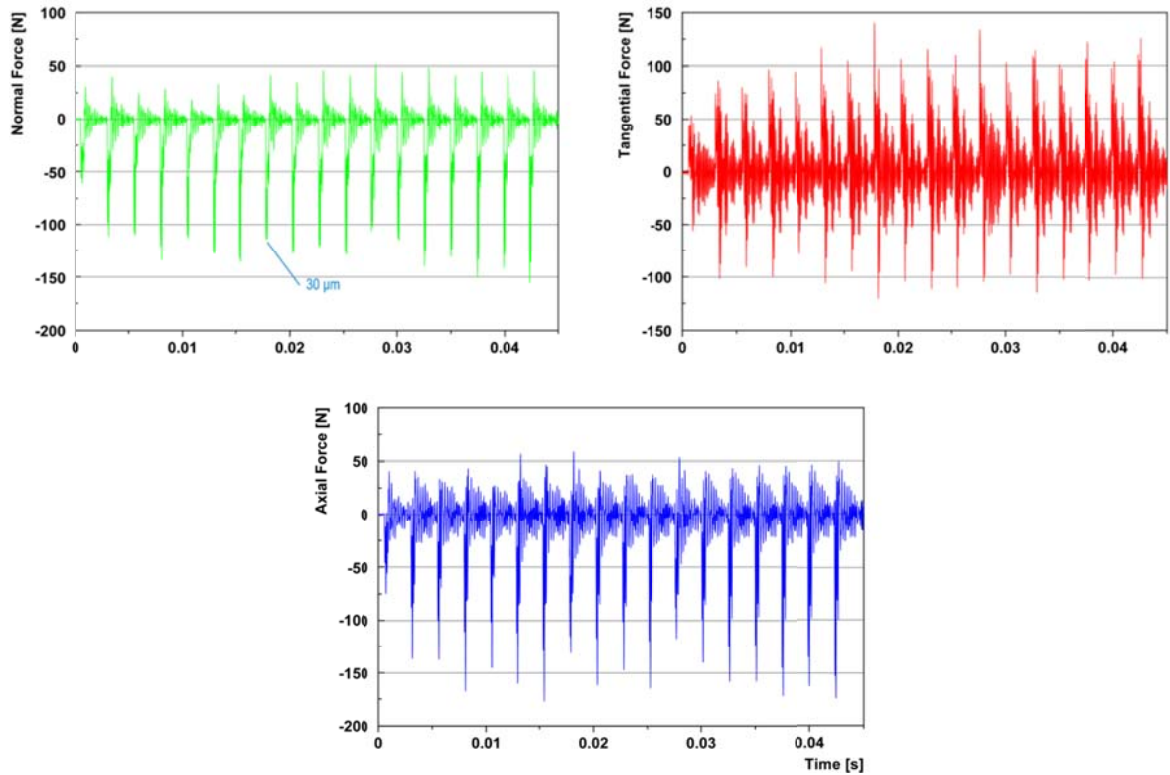

Fig. 13: Forces in all three directions for the whole process and all engagements are shown.

The oscillations in the force results are due to the impacts of the diamond grains with the two rotating workpieces. Normal, tangential and axial forces are 115 N, 141 N and 132 N respectively at 30 μm depth of cut.

## 4. Conclusion

In daily-bases industrial design of the cutting tools, one must consider the flash temperature which leads to wear of the tool and changing the form of the cutting geometry. Special form of the cutting geometry in the design phase will also lead to lower flash temperature. However since its measurement is hard, normally it is approximated. Thus, cutting tool manufactures, as well as the users of the cutting tools are the potential applicants. A novel test setup is developed to measure flash temperatures and forces simultaneously for grinding applications. For measuring the flash temperature an innovative method to measure and analyze the temperature through the diamond grain in the cutting zone by two-color pyrometer are proposed. . The developed tool is analyzed based on FEM to guarantee stability under high rotational velocities. The frequency behavior of the tool is simulated, to find out the most critical eigenmodes. After performing the harmonic response analysis, the cutting velocity of 65 m/s is considered. This led to not exciting the critical eigenmodes of the system. Additional dynamic balance is performed to overcome the manufacturing uncertainty. However, it is shown that some vibrations during the machining exist, which arise from other resources such as dynamics of the complete system. This could be also interpreted due to elastic deformation of the tool-holder, which was manufactured out of aluminum to reduce the weight on the dynamometer. It is suggested that in future, light weighted tool-holders with high strength from materials such as Titanium alloys will be used. Therefore, quasi-static, dynamic, modal and harmonic response analyses can be used as a guide line for design of the special tools and perform high precision interaction of the tool-material.

The concept of measuring temperature at the cutting edge through diamond grain is tested successfully. Measurement results show that with help of this accurate setup and two-color pyrometer, the flash temperature can be measured accurately in the interaction zone. The flash temperature of 1380 °C is detected at 30 μm depth of cut over scratch length of 20.1 mm. After performing EDX analysis, it is found that Titanium in form of buildup-edge is formed on the cutting edge of the diamond. Since the Titanium has the melting temperature in the same region as the measured temperature, the formation of buildup-edge in the cutting edge can be explained. Therefore, for grinding of titanium alloy these process parameters due to high flash temperature and formation of buildup-edge must not be considered and lower depth of cut or lower feed and cutting speed must be taken into account. Analysis of eigenfrequencies is made in order to avoid vibrations of the rotating workpiece. It is shown, that the chosen rotation velocity is far from the first eigenfrequency of the rotating workpiece. Moreover the assumed forces are quasi-static forces and not for the measured peaks of dynamic forces. Furthermore, the forces are measured on the grain and not on the workpiece, therefore for correct comparison, the eigenfrequencies of the measurement device and not of the workpiece have to be analyzed. The force measurement is challenging because of the high process dynamics. Oscillation occurs after the first inlet impulse and does not decline to zero before the next scratch takes place. Force peaks of normal, tangential and axial of 115 N, 141 N and 132 N respectively are evaluated at 30 μm depth of cut. The results of this work open new possibilities for flash temperature measurements, machining process parameters optimizations of Ti6Al4V and the advantage of finite element analysis in design of special tool holders as it is described.

**Acknowledgements** The authors would like to thank MTTRF for their support for 5-axis machining center of Mori-Seiki NMV5000 DCG-MTTRF, support from Fredy Kuster, Marco Boccadoro, Sandro Wigger, Daniel Spescha, Lukas Seeholzer and Michael Steck.